\documentclass[11pt,preprint2]{emulateapj}

\usepackage[normalem]{ulem}

\shorttitle{HCO$^+$ and HCN (3-2) absorption toward Cen A}
\shortauthors{Muller \& Dinh-V-Trung 2009}

\begin{document}

\title{HCO$^+$ and HCN J=3-2 absorption toward the center of Centaurus A}

\author{S\'ebastien Muller \altaffilmark{1}
\& Dinh-V-Trung \altaffilmark{1,2}
}

\altaffiltext{1}{Academia Sinica, Institute of Astronomy and Astrophysics (ASIAA),
P.O. Box 23-141, Taipei 106, Taiwan}
\altaffiltext{2}{On leave from Center for Quantum Electronics, Institute of Physics, Vietnamese Academy
of Science and Technology, 10 DaoTan, ThuLe, BaDinh, Hanoi, Vietnam.}

\begin{abstract}

We have investigated the presence of dense gas toward the radio source Cen A by looking at the
absorption of the HCO$^+$ and HCN (3-2) lines in front of the bright continuum source
with the Submillimeter Array. We detect narrow HCO$^+$ (3-2) absorption, and tentatively HCN (3-2),
close to the systemic velocity. 
For both molecules, the $J=3-2$ absorption is much weaker than for the $J=1-0$ line.
From simple excitation analysis, we conclude that the gas density is on the order of a few $10^4$
cm$^{-3}$ for a column density $N$(HCO$^+$)/$\Delta V$ of $3 \times 10^{12}$ cm$^{-2}$~km$^{-1}$~s
and a kinetic temperature of 10 K. In particular, we find no evidence for molecular gas density
higher than a few $10^4$ cm$^{-3}$ on the line of sight to the continuum source. We discuss the
implications of our finding on the nature of the molecular gas responsible for the absorption 
toward Cen A.

\end{abstract}


\section{Introduction}

The giant elliptical galaxy NGC~5128 hosts the powerful central radio continuum source Cen~A
and contains significant amount of gas and dust projected in front of it. It is therefore a
good target for absorption study. It was the first radio galaxy in which absorption
in neutral hydrogen was detected (\citealt{rob70}). Since then, numerous molecules have been
detected in absorption toward Cen A as well, first at radio wavelengths, such as H$_2$CO
(\citealt{gar76}), C$_3$H$_2$ (\citealt{sea86}), OH, NH$_3$ (\citealt{sea90}), and further at
millimeter wavelengths with the Swedish-ESO Submillimeter Telescope (SEST): CO, HCO$^+$, HCN,
HNC, CN, CS (\citealt{isr90, eck90,wik97}). The molecular absorption profile is best revealed
from the HCO$^+$(1-0) spectrum published by \cite{wik97}, owing to the large opacity of the
transition, and to the high signal to noise and good spectral resolution of the data.

The HCO$^+$(1-0) absorption profile can be mainly decomposed into two components: {\em i)} a
series of narrow ($1-2$ km~s$^{-1}$ wide) lines with absorption depth ranging from $\sim 50$\%
to nearly 100\% of the continuum intensity, located in the heliocentric velocity range between
540 and 560 km~s$^{-1}$, i.e. close to the systemic velocity, and {\em ii)} a broad absorption
feature between 560 and 640 km~s$^{-1}$, of $\sim 10$\% of the continuum, also including several
narrow components (10 to 50\% of the continuum). Following \cite{wik97}, we will refer to these
components as Low Velocity Complex (LVC) and High Velocity Complex (HVC), respectively.

Similarly, the HI absorption profile shows numerous narrow components, although some with no
molecular counterparts, and vice versa. The estimates of the systemic velocity range between
$V_{\rm HEL} = 536$ and 551 km~s$^{-1}$ in the literature (see e.g. the review by \citealt{isr98}).
The existence of redshifted HI absorption, seen only in front of the radio core and not against
the inner lobes, has been interpreted by \cite{vanhul83} as gas infall toward the central source.
Eventually, this gas infall could be sufficient to fuel a supermassive black hole (\citealt{vangor89}).
\cite{sar02} showed that a weak and broad HI absorption, previously undetected and corresponding
in velocity to the molecular HVC, is occuring only in front of the nucleus. The region showing
redshifted gas is limited in size to $\leq 100$ pc from the radio core, and is thus a strong
candidate for a circumnuclear disk. New HI observations with the Australia Telescope Compact Array
(ATCA) allowed \cite{mor08} to also detect blueshifted absorption, thus favoring the interpretation
in terms of a rotating circumnuclear disk. It is not clear, however, what is the illuminating
background continuum source at this frequency. Indeed, \cite{tin01} show that, between 2 and 5 GHz
the innermost part of the radio continuum source is affected by free-free absorption that might be
caused by circumnuclear ionised gas. The absorption of hard X-ray, indicating a column density of
$10^{23}$ atoms cm$^{-2}$ of absorbing gas in front of the central source (\citealt{eva04}), also
suggests the presence of circumnuclear material around the supermassive black hole, as expected from
the AGN unification models (\citealt{ant93}).

The central region of Cen A is heavily obscured in the optical, with A$_{\rm V} \sim 15$
(\citealt{isr90,eck90}). The presence of a circumnuclear molecular disk in Cen A, with
extent of about $100 - 200$ pc in radius, has been inferred by several authors
(\citealt{isr90,haw93,ryd93,lis01}) on the basis of infrared and submillimeter emission
data. Recently, \cite{neu07} mapped the H$_2$ 2 $\mu$m lines emission at $\sim 0.1\arcsec$
resolution toward the central region with a field of view of $3\arcsec \times 3\arcsec$
($\sim 50 \times 50$ pc). Their best fit of the H$_2$ velocity field, using a warped
disk model, indicates a median disk inclination angle of $45\degr \pm 12\degr$ and a
black hole mass of $\sim 5 \times 10^7$ M$_\odot$.

The HVC absorption is significantly redshifted with respect to the systemic velocity, and 
spread over a larger velocity range than that of the LVC, forming, as seen in
the HCO$^+$ (1-0) line, a broad and continuous absorption over more than 50 km~s$^{-1}$ wide.
For these reasons, \cite{wik97} suggest that the HVC absorption components could originate in the
circumnuclear disk. The LVC components, on the other hand, may correspond to intervening gas in
the galactic disk of Cen A, although it is difficult to determine their locations. 
Alternatively, \cite{eck99} proposed that the general velocity structure of the absorption could be
explained kinematically with a tilted-ring model and high-altitude clouds, not necessarily
requiring the presence of molecular gas close to the active nucleus.

In any case, the physical conditions of the absorbing molecular gas are still poorly known. Previous
analysis of CO multi-transition single-dish observations were conducted (\citealt{isr91}), but suffer
from strong contamination by line emission. The situation is better for $^{13}$CO, as the line emission
is greatly reduced. From analysis of the $^{13}$CO (1-0) and (2-1) line absorption, \cite{eck90} derived
volume density $n$(H$_2$) of a few times $10^4$ cm$^{-3}$ for the narrow components in the LVC.
Van \cite{vanlan95} observed the ground state main lines of OH (rest frequencies of 1665 and 1667 MHz) with
the ATCA and estimate similar gas density for LVC absorption components. Interestingly, they also
observed the two satellite lines at 1612 and 1720 MHz and find that these lines have a strong conjugate
behavior, one in absorption, and the other equally strong in emission. This is particularly remarkable
in the HVC, because the corresponding absorption is weak in the main OH lines. Van \cite{vanlan95}
note that such a behavior could occur at high density $n \gtrsim 10^6$ cm$^{-3}$.
To the best of our knowledge, no other molecule has been observed in different rotational
transitions, and the physical properties corresponding to the HVC components are mostly unknown.

With the aim to study the properties of the gas in front of the nuclear continuum source, and
particularly the dense gas component as a probe of the potential circumnuclear disk, we have
observed the high density gas tracers HCO$^+$ and HCN (3-2) transitions toward the center of
Centaurus A with the Submillimeter Array. We present our observations and results in \S\ref{obs}
and \S\ref{results}. Implications are discussed in \S\ref{discuss}. A thorough review on Cen A
is given by \cite{isr98}.

\section{Observations and data reduction} \label{obs}

The $J=3-2$ transitions of HCO$^+$ (267.5576 GHz) and HCN (265.8864 GHz) were observed toward Cen A
with the Submillimeter Array (SMA), on 2007 April 7th. The array was composed of eight antennas in a
compact north-south configuration, optimized for observations of sources in the southern hemisphere.
The projected baselines ranged between 5 and 110 m over the course of the observations. The phase
reference was set to the position of Cen A at
(R.A., Dec.)$_{\rm J2000}$ = (13$^{\rm h}$25$^{\rm m}$27$\fs$60, $-43\degr$01$\arcmin$09$\farcs$0).

We observed Cen A for a total on-source integration time of 4.3 hours. The zenith atmospheric opacity
at 225 GHz was between 0.1 and 0.15. System temperatures ranged between 200 and 600 K for the different
antennas, except for one antenna, for which the system temperature was between 600 and 900 K.
The data were calibrated with the software package MIR/IDL. The bright radio sources 3C273, 3C279 and
Ganymede were observed for bandpass calibration, for which an antenna-based solution was adopted.

The heterodyne receivers were tuned to observe simultaneously the HCO$^+$ and HCN (3-2) transitions,
both placed in the lower sideband (LSB). The $\sim 2$ GHz bandwidth
of the LSB was divided into 24 spectral windows (``chunks''), 104 MHz ($\sim 95$ km~s$^{-1}$)
wide each, and slightly overlapping in frequency.

The continuum emission (Fig.\ref{map}a) was reconstructed by averaging all the LSB line-free chunks, resulting in a
total spectral bandwidth of 1.6 GHz. The absorption components, especially around $V_{\rm HEL} \sim
550$ km~s$^{-1}$, have narrow linewidths ($\sim$ 1-5 km~s$^{-1}$). We therefore used a velocity
resolution of $\sim$ 0.2 km~s$^{-1}$ (i.e., 512 channels/chunk) for the chunks corresponding to the
HCO$^+$ and HCN lines. All the other chunks were set to a spectral resolution of 0.8125 MHz (or a
velocity resolution of $\sim 0.9$ km~s$^{-1}$). To further improve the signal to noise ratio, the
spectral resolution of all chunks has been smoothed to 1.625 MHz (i.e., $\sim 1.8$ km~s$^{-1}$). 

We used the strong and unresolved continuum emission to self-calibrate the line visibilities. We estimate
the continuum flux density of Cen A to be $\sim 6.6$ Jy at 266 GHz, with uncertainty of order of 20\%,
by using Ganymede as a flux calibrator. The complex gains (amplitude and phase {\em vs} time) were 
self-calibrated on the continuum visibilities. The amplitude was normalized to the continuum level.

Continuum-subtracted channel maps were produced for both lines. The deconvolved maps, integrated
over the LVC (from 535 to 560 km~s$^{-1}$) and HVC (from 565 to 620 km~s$^{-1}$) velocity ranges,
are shown in Fig.\ref{map}c--f. Adopting natural weighting, the uv-coverage of our SMA observations
yields a synthesized beam of $4.0\arcsec \times 2.6\arcsec$, with a position angle of $9\degr$.
A fit of the calibrated line visibilities, with the GILDAS/MAPPING task UVFITS, using a point source model
with fixed position (at the phase center) but free amplitude, resulted in the spectra shown in Fig.\ref{spec}.



\section{Results} \label{results}

\begin{figure} \includegraphics[width=8cm]{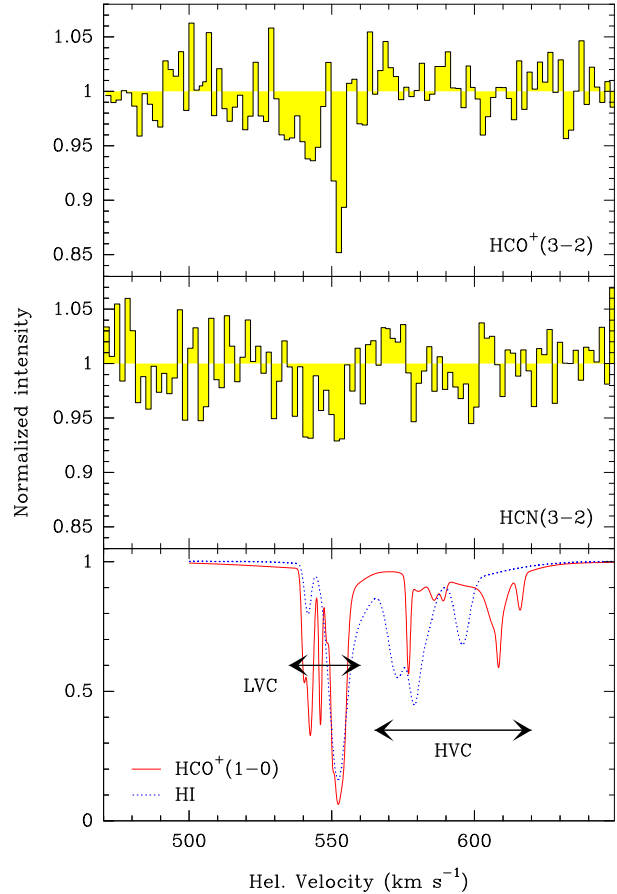}
\caption{Spectra of the HCO$^+$ ({\em top}) and HCN ({\em middle}) (3-2) absorption lines toward
Cen A. The velocity resolution is $\sim 1.8$ km~s$^{-1}$. The profile of the HCO$^+$ (1-0) ({\em full line},
observed by \citealt{wik97}) and HI absorptions ({\em dotted}, obtained by \citealt{sar02}) are shown in
the bottom frame. The velocity range of the LVC and HVC is also indicated.}
\label{spec} \end{figure}

The HCO$^+$ and HCN (3-2) absorption spectra toward Cen A, obtained with the SMA, are presented in
Fig.\ref{spec}. For comparison, we also include the absorption profiles of HCO$^+$ (1-0) and HI,
as parametrized by \cite{wik97} and \cite{sar02}, respectively.

\begin{figure} \includegraphics[width=8cm]{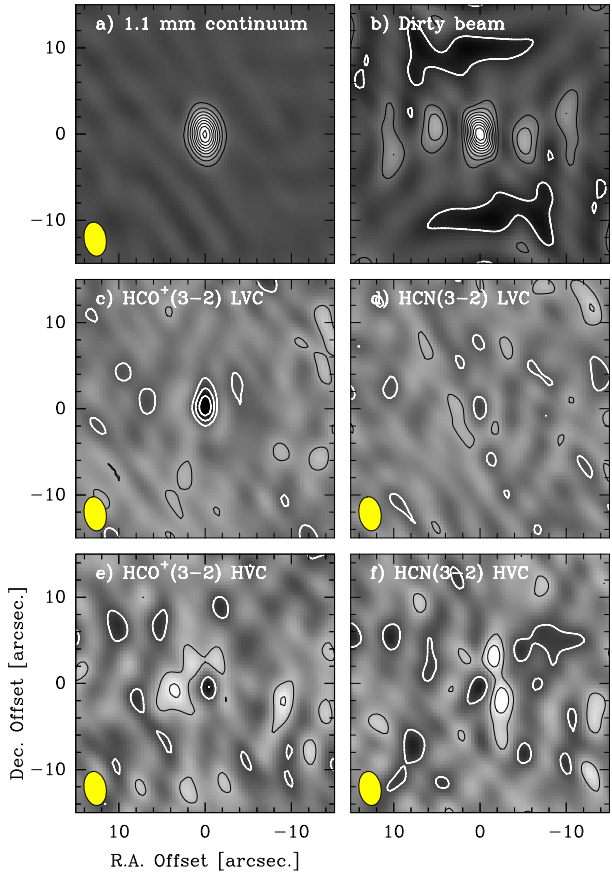}
\caption{a) 1.1 mm normalized continuum emission. Contour levels are drawn every 10\% ($\sim 23\sigma$);
b) Dirty beam corresponding to the uv-coverage of our SMA observations (contour levels every 10\%,
in black for positive contours and white for negative ones);
c,d,e,f) Maps of the HCO$^+$ and HCN(3-2) line intensity integrated over the LVC and HVC velocity ranges.
The continuum emission has been subtracted from visibilities before imaging. 
Contour levels are shown every 2$\sigma$, i.e., 1.2 Jy~beam$^{-1}$~km~s$^{-1}$ for LVC maps and
1.8 Jy~beam$^{-1}$~km~s$^{-1}$ for HVC maps, in black for positive contours and white for negative ones.
The synthesized beam ($4.0\arcsec \times 2.6\arcsec$,
P.A. $= 9\degr$) is shown at the bottom left corner of each map.}
\label{map} \end{figure}

The HCO$^+$ (3-2) absorption is clearly detected in the LVC velocity range (see also Fig.\ref{map}c).
A strong and narrow ($\sim 1$ km~s$^{-1}$ wide) line,
located at $\sim 552$ km~s$^{-1}$, matches with the deepest absorption component seen in the HCO$^+$ (1-0)
spectrum. The absorption depth reaches 16\% of the continuum level. The r.m.s. noise level, measured over
line-free channels, is 2.5\% of the continuum. Blueward of this line, at a velocity of 543 km~s$^{-1}$,
a second weaker component can be identified and probably results from the blend of several narrow ($\leq
2$ km~s$^{-1}$) line components, identified in the high velocity resolution spectrum of \cite{wik97}.
The signal to noise ratio of this feature, however, is limited, and barely reaches 3$\sigma$ at the peak. 

Concerning the HCN (3-2) line, we tentatively detect counterparts to the 543 and 552 km~s$^{-1}$ components,
although the signal to noise ratio is poor, with maximum absorption of $\sim 7$\%. We note that the hyperfine
structure of the HCN (3-2) transition concentrates about 93\% of the total line intensity within 0.3 MHz
(e.g., \citealt{mak74}), and is therefore not resolved given our spectral resolution.

Given the rms noise level of our observations, $\sim 2.5$\% of the continuum level, we do not detect
any other absorption feature. Especially, no counterparts of the HVC components are detected.
Also, no HCO$^+$ or HCN (3-2) {\em emission} is evident.


\section{Discussion} \label{discuss}

\subsection{Continuum background and filling factor}

The 266 GHz continuum emission remains unresolved at our angular resolution of $\sim 3\arcsec$. From
Very Long Baseline Interferometry (VLBI), the nuclear region of Cen A is shown to harbor a core-jet
radio structure (\citealt{tin98}). As the jet has a steep radio spectrum, with negligible flux at
millimeter wavelengths, the molecular absorption should however occur toward the radio core.
\cite{kel97} measured the size of the radio core as $0.5 \pm 0.1$ mas in diameter from 43 GHz VLBI
observations, corresponding to a linear dimension of $\sim 0.01$ pc, and yielding a brightness
temperature on the order of $10^{10}$ K. The temperature of any molecular gas in the line of sight
is therefore expected to be completely negligible with respect to that of the continuum. Given the
very small apparent size of the continuum source at millimeter wavelengths, we expect the continuum
source to be completely covered by the absorbing clouds, and will assume a filling factor of unity in
our analysis.
This assumption is supported by VLBA data of the OH 18 cm absorption, which appears to cover a large
fraction the continuum source (van \citealt{vanlan05}).

In the following, we will compare our HCO$^+$ and HCN (3-2) data to previous observations of the
corresponding (1-0) transitions obtained in 1995-1996 by \cite{wik97}. To the best of our knowledge,
no other observations of HCO$^+$ or HCN absorption toward Cen A were attempted inbetween. The
comparison might thus be affected by time variations, although \cite{wik97} could not detect
significant changes of more than 10\% between their HCO$^+$ (1-0) absorption profile and that
observed 7 years before by \cite{eck90}.



\subsection{Excitation analysis}

In order to study the excitation conditions of the HCO$^+$ and HCN (3-2) lines in the line of
sight to Cen A, we have used a molecular excitation code based on the large velocity gradient
(LVG) approximation. Molecular data for HCO$^+$ and HCN, including the energy levels, the
radiative transition rates and collisional cross sections, were taken from the Leiden Atomic
and Molecular Database (\citealt{sch05}). In our calculations, we include all rotational levels
up to the $J=10$ level. The temperature of the cosmic background radiation field is set to
2.73 K. Our code need three basic parameters as inputs: the molecular hydrogen density $n$(H$_2$),
the kinetic temperature $T_{\rm kin}$ and the gas column density per unit velocity $N$/$\Delta V$.
The statistical equilibrium equations setup in the framework of the LVG approximation are solved
iteratively using a Newton-Raphson method. We consider that convergence is achieved when the
relative change in level populations is less than $10^{-3}$ between two successive iterations.

\begin{figure} \includegraphics[width=7cm]{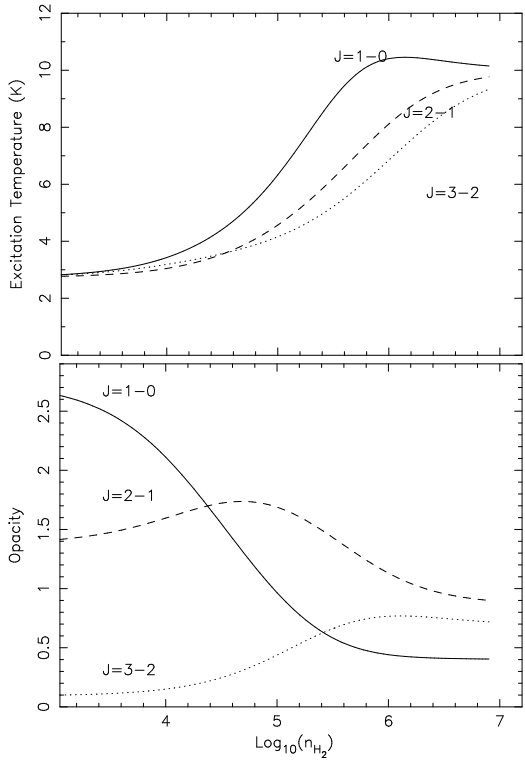}
\caption{Excitation temperature and opacity of the $J=1-0$, $2-1$ and $3-2$ transitions of HCO$^+$ 
as a function of gas density in the case $T_{\rm kin} = 10$ K and $N$(HCO$^+$)/$\Delta V = 3 \times
10^{12}$ cm$^{-2}$~km$^{-1}$~s.}
\label{hco} \end{figure}

We present the results of our calculations in Fig.\ref{hco} and \ref{hcn}, where the changes
of excitation temperature and line opacity as a function of molecular hydrogen density are shown
for the $J= 1-0$, $J=2-1$ and $J=3-2$ transitions of HCO$^+$ and HCN molecules, respectively.
We assume a kinetic temperature of 10 K, within the range of temperatures commonly inferred in
previous studies (\citealt{isr91,eck90}). The column density per unit velocity is set to
$3 \times 10^{12}$ cm$^{-2}$~km$^{-1}$~s for HCO$^+$, and $2 \times 10^{12}$ cm$^{-2}$~km$^{-1}$~s
for HCN, respectively. These values are consistent with that derived by \cite{wik97}.

At low gas density, i.e., $n$(H$_2$) $\sim 10^3$ cm$^{-3}$, the HCO$^+$ molecules are not excited
to high $J$ levels. As can be seen from Fig.\ref{hco}, the excitation temperature is low for
all transitions shown. Most of the HCO$^+$ molecules stay in the $J=0$ and $J=1$ levels, resulting
in high opacity for the $J=1-0$ and $2-1$ transitions. As the gas density increases, more HCO$^+$
molecules are excited to higher $J$ levels. The opacity of the $J=1-0$ transition drops rapidly,
while the opacity of the $J=3-2$ transition increases at nearly the same pace. We note here the
non-intuitive behaviour of the $J=2-1$ transition: the opacity of this transition increases slightly
with density, peaking at a hydrogen density of $10^5$ cm$^{-3}$ and then falls off at higher density.
The $J=2-1$ line opacity remains optically thick for the whole range of density considered in our
calculations. When the gas density reaches a few $10^5$ cm$^{-3}$, the opacity of both the $J=1-0$
and $J=3-2$ lines becomes comparable. At high gas density, greater than a few times $10^6$ cm$^{-3}$,
all the three lines are thermalized. In this case, the opacity of the lines are determined by the
usual Boltzmann distribution of level population. As we can see from Fig.\ref{hco}, the $J=3-2$ line
has a significantly higher opacity than the $J=1-0$ line. From the behaviour of the line opacity as
a function of gas density, we suggest that the comparison of opacity for the $J=1-0$ and $J=3-2$ pair
constitutes a sensitive constraint to the gas density.

The excitation temperature and line opacity for HCN molecules behave in a similar way (Fig.\ref{hcn}).
However, the HCN lines become thermalized at higher density, above $10^7$ cm$^{-3}$, due to lower
collision rates in comparison to that of HCO$^+$.

\begin{figure} \includegraphics[width=7cm]{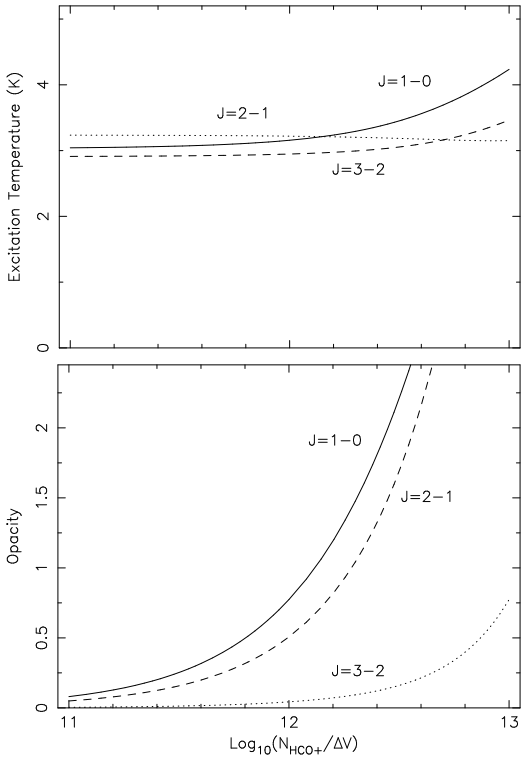}
\caption{Excitation temperature and opacity for the $J=1-0$, $2-1$ and $3-2$ transitions of HCO$^+$
for a gas density of $n$(H$_2) = 10^4$ cm$^{-3}$ and a kinetic temperature of $T_{\rm kin} = 10$ K.}
\label{ncol} \end{figure}

For completeness, we also calculate the excitation temperature and opacity as a function of the column
density of HCO$^+$ for a gas density of $n$(H$_2) = 10^4$ cm$^{-3}$ and a kinetic temperature of 10 K
(see Fig.\ref{ncol}). The opacity of the three transitions increases monotonically with the HCO$^+$
column density. Especially, the HCO$^+$ (3-2) transition shows appreciable opacity ($\tau \gtrsim 0.1$)
only above $N$(HCO$^+$)/$\Delta V$ $> 10^{12}$ cm$^{-2}$~km$^{-1}$~s, where the (1-0) and (2-1)
transitions become rapidly optically thick. The comparison of the opacity for the $J=1-0$ and $3-2$
transitions of HCO$^+$ thus gives useful informations on the column density of absorbing gas.

\subsection{IR excitation}

So far, we have neglected the effect of the infrared radiation field emitted by dust particles,
on the excitation of the HCN and HCO$^+$ molecules. Both molecules can be excited from the ground
state $J=0$ to the $J=1$ rotational level of the first bending mode $\nu_2=1$ by the absorption
of IR photons at 14 $\mu$m for HCN and 12 $\mu$m for HCO$^+$. Subsequent decay transfers a
fraction of the excited molecules to the $J=3$ level of the ground state.


The strength of the local radiation field at the location where the absorption arises is still poorly
known. Van \cite{vanlan95} adopt a representative radiation field consisting of two components: a warm dust
component at a temperature of 150 K and a cooler dust component at 43 K. Obviously, only the warm dust
component contributes to the IR radiation at the absorption wavelengths of HCN and HCO$^+$ molecules.

\begin{figure} \includegraphics[width=7cm]{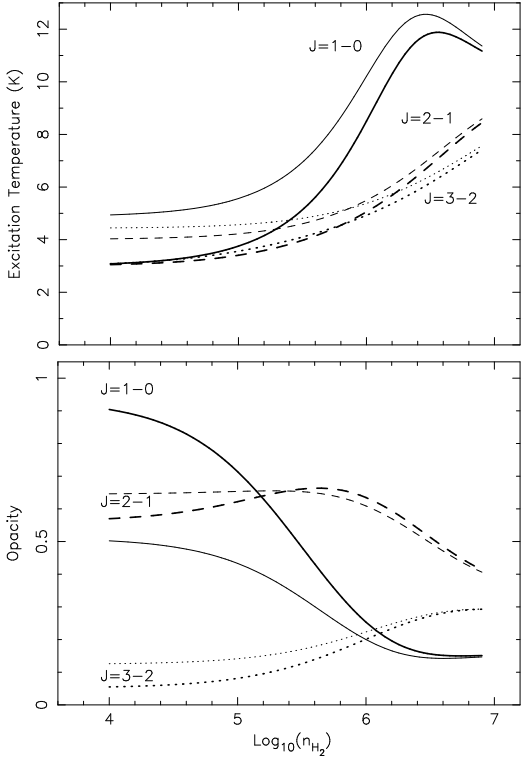}
\caption{Excitation temperature and opacity of the $J=1-0$, $2-1$ and $3-2$ transitions of HCN 
as a function of gas density in the case $T_{\rm kin} = 10$ K and $N$(HCN)/$\Delta V = 2 \times 
10^{12}$ cm$^{-2}$~km$^{-1}$~s. Thick curves correspond to the case of excitation by collision
only, thin curves with IR pumping included. The parameters of the radiation field are taken from
van \cite{vanlan95}: T$_D=150$ K, $\Delta \Omega =0.1$, E$(B-V)=1$ and $\beta=1$.}
\label{hcn} \end{figure}

We have repeated our LVG calculations for the HCN molecule taking
directly into account the IR pumping. The formulae for the excitation and de-excitation rates are taken
from \cite{deg84}. Results are shown in Figure \ref{hcn}.
Because of IR pumping, the excitation temperatures of all rotational transitions increase noticeably
in comparison to the case without IR pumping for gas densities below 10$^6$ cm$^{-3}$.
At higher densities approaching 10$^7$ cm$^{-3}$, the effect of IR pumping is very small, as expected.
We note that at low gas densities, where more molecules are excited by IR pumping to higher $J$ levels,
there is an increase in the opacity of the $J=3-2$ transition while the opacity of $J=1-0$ transition
is reduced. Interestingly, the opacity of $J=2-1$ transition remains approximately the same as in the
case without IR pumping. Because both HCN and HCO$^+$ molecules have similar transition probability
(\citealt{gar06})
and vibrational transition frequency, we expect very similar results for the excitation of HCO$^+$
molecules. It is clear from Figure \ref{hcn} that with or without the IR pumping the behavior of the line
opacity as a function of the gas density does not change qualitatively. Observationally, the $J=3-2$
absorption of HCO$^+$ and HCN is very weak, suggesting that the IR pumping probably does not play an
important role in the excitation of these molecules. Therefore our conclusion in
the previous section is not affected by IR pumping.

\subsection{Physical conditions for the LVC components}

The LVC components, i.e., between $V_{\rm HEL} = 535$ and 560 km~s$^{-1}$, all exhibit strong
absorption in the HCO$^+$ (1-0) transition with opacities $\gtrsim 1$, whereas our SMA data
shows that the HCO$^+$ (3-2) absorption of the same features is relatively weak, with opacities
$\lesssim 0.2$. From Fig.\ref{hco}, we can thus estimate that, in order to be consistent with the
observations, the density of the absorbing gas should be about $10^4$ cm$^{-3}$ for a HCO$^+$
column density per velocity unit on the order of $3 \times 10^{12}$ cm$^{-2}$~km$^{-1}$~s.

Similarly, the HCN (1-0) absorption components reach an opacity of about 0.9, while our tentative
detection of HCN (3-2) absorption suggests opacity well below 0.1. This is also consistent with
gas density of about $10^4$ cm$^{-3}$ and column density per velocity unit of $\sim 2 \times
10^{12}$ cm$^{-2}$~km$^{-1}$~s (Fig.\ref{hcn}).

Our results are therefore comparable to that obtained by \cite{eck90} from analysis of the $^{12}$CO and
$^{13}$CO (1-0) and (2-1) lines for the deepest absorption feature at $V_{\rm HEL} = 552$
km~s$^{-1}$. They estimate a kinetic temperature of no more than 10 K and a gas density of around
$2 \times 10^4$ cm$^{-3}$. They also note, however, that the hyperfine structure line ratios for the
HCN and CN (1-0) transitions could suggest a clumpy medium, with density possibly up to $10^6$ cm$^{-3}$.
We do not see evidence of such high density medium from our data.

\subsection{Constraints for the HVC components}

The main goal of these observations was to probe the physical properties of the gas associated with the HVC
absorption, possibly associated with the circumnuclear disk around the supermassive black hole in Cen A.
The HVC absorption is best detected in HCO$^+$ (1-0), but also appears in HI (although with no one to one
correspondence, see e.g. \citealt{sar02}), OH ground state lines (van \citealt{vanlan95}), HCN and HNC (1-0), and
CS (2-1) lines (\citealt{wik97}).

According to the results of our excitation analysis, the non-detection of HCO$^+$ (3-2) absorption counterpart
to HCO$^+$ (1-0) components suggests that the gas density is lower than a few times $10^4$ cm$^{-3}$. 

\cite{eck99} proposed a kinematical explanation for the HVC absorption, which could be
caused by clouds located at large galactocentric radii on the order of 0.5 kpc and high altitude of
$\sim 300$ pc above the disk. Whereas our observations give contraints on the physical conditions of
the molecular gas, particularly on density, they alone can not rule out the existence of a circumnuclear
disk neither confirm the kinematic interpretation of the complex absorption system.

\section{Conclusions}

We have observed the absorption from the HCO$^+$ and HCN (3-2) transition toward the center of Centaurus A
with the Submillimeter Array. At least two absorption components are identified in the HCO$^+$ (3-2)
spectrum. A first narrow component, located at $V_{\rm HEL} \sim 552$ km~s$^{-1}$, reaches a depth of
about 16\% of the continuum level and corresponds to the deepest absorption component observed in other
molecules and HI. A second weak component, approximately 10 km~s$^{-1}$ blueward and
7\% deep, probably corresponds to the blend of several narrow components identified in the HCO$^+$ (1-0)
spectrum by \cite{wik97}. Given our sensitivity of 2.5\% of the continuum level, no counterparts of the
redshifted absorption components seen in HCO$^+$ (1-0) between 570 and 620 km~s$^{-1}$ are detected.
Absorption from the HCN (3-2) transition is tentatively detected around $V_{\rm HEL} \sim 552$ km~s$^{-1}$.

While the sensitivity and dynamic range of our SMA observations are limited, the weak absorption 
in the HCO$^+$ and HCN (3-2) transitions, as compared to the corresponding $J=1-0$ absorption,
provides useful informations about the physical conditions of the absorbing gas.
We have performed a simple excitation analysis for the $J=1-0$, $2-1$ and $3-2$ transitions of HCO$^+$
and HCN molecules. We find that the HCO$^+$ (1-0) and (3-2) pair of transitions is an excellent indicator
of the absorbing gas density. Absorption components close to the systemic velocity (LVC) have density on
the order of a few times $10^4$ cm$^{-3}$, for a column density of a few times $10^{12}$ cm$^{-2}$.
The non-detection of absorption counterparts to the HVC redshifted components suggests corresponding
density of $10^4$ cm$^{-3}$ or lower. Thus, nowhere on the line of sight to the central continuum source
of Cen A is to be found molecular gas with density higher than a few $10^4$ cm$^{-3}$.
The inclusion of IR excitation in our model does not change substantially these results.
Either the line of sight to the radio continuum source does not intersect the circumnuclear disk,
and the different absorption components arise at different locii in the galactic disk, or
the density of the absorbing gas in the circumnuclear disk is lower than a few $10^4$ cm$^{-3}$.

\acknowledgements
We thank the SMA staff for their very competent assistance with these observations.
We are greatful to the anonynous referee for providing us with constructive comments.
The Submillimeter Array is a joint project between the Smithsonian Astrophysical Observatory
and the Academia Sinica Institute of Astronomy and Astrophysics and is funded by the Smithsonian
Institution and the Academia Sinica.

\facility{{\it Facility:} \facility{Submillimeter Array}}

\end{document}